\documentstyle[aps,prl,multicol,epsf]{revtex}
\begin{document}

\title{Time distribution and loss of scaling in granular
flow\thanks{This paper is dedicated to Professor
Franz Schwabl on the occasion of his 60th birthday}}
\author{Bosiljka Tadi\'c}

\address{
Jo\v{z}ef Stefan Institute,
P.O. Box 3000, 1001-Ljubljana, Slovenia }

%\date\today

\maketitle
\begin{abstract}
Two cellular automata models with directed mass flow and internal time scales
are studied by numerical simulations. Relaxation rules are a combination of
probabilistic critical height (probability of toppling $p$) and
deterministic critical slope processes with internal correlation time $t_c$
equal to the avalanche lifetime, in Model A, and  $t_c\equiv 1$,
in Model B. In both cases nonuniversal scaling properties of avalanche
distributions are found for $p\ge p^\star $, where $p^\star $ is related
to directed percolation threshold in $d=3$.
Distributions of avalanche durations for $p\ge p^\star $ are studied in
detail, exhibiting multifractal scaling behavior in model A, and finite
size scaling behavior in model B, and scaling exponents are determined
as a function of $p$.
At $p=p^\star $ a phase transition to noncritical steady state occurs.
Due to difference in the  relaxation mechanisms, avalanche statistics at
$p^\star$ approaches the parity conserving universality class in Model A,
and the  mean-field universality class in Model B. We also estimate
 roughness exponent at the transition.
\end{abstract}
%\narrowtext
\begin{multicols}{2}

{\section{Introduction}}

Dynamics of granular materials represents an important practical and
theoretical problem. A new theoretical approach
to the problem of driven granular flow has been initiated in the past
few years \cite{Bak_book}, which is  motivated by the observed scaling
 behavior
both in the laboratory granular piles and in natural landslides
\cite{Jaegeretal,Heldetal,RVK,Bretzetal,rice,Noever,landslides}.
It has been recognized that
the collective dynamics of grains may lead to a self-organized critical
(SOC) states \cite{Bak_book}, characterized by scaling
 properties of sandslides (avalanches).
 Moreover, dynamics may depend on various parameters, such as
dimension and mass of individual grains and quality of their contact
surfaces, and on the external conditions. By varying some of these
parameters in a controlled manner, steady states with different
characteristics are reached, and a phase transition to a
steady state with no long-range correlations occurs when a parameter
is  varied through certain  critical value \cite{rice}.

Various  cellular automata models have been introduced so far
to mimic stochastic variations  in the conditions of toppling
\cite{rice-transport,CAM-gran,CAM-gran1,Lauritsen,LTU,BT-gran}.
One-dimensional
rice-pile automata with stochastic critical slope rules have been useful
in  understanding  transport properties of rice piles \cite{rice-transport}.
Relaxation rules in these models are a kind of branching processes
with internal stochasticity. In two dimensions, two models studied
in Refs.\ \cite{LTU,BT-gran} utilize mixed dynamic critical slope (CS)
and critical height (CH) rules, motivated by the observed nonuniversality
of the emergent critical states in natural landslides (for a recent review
see Refs.\ \cite{Turcotte,erosion}).

In the present work we extend the study of the models of Refs.\
\cite{BT-gran,LTU}, which we term model A and B, respectively.
 In these models
stochastic toppling by the CH mechanism is controlled by an external
parameter---probability of toppling $p$, which can be attributed to
variations in the external conditions (e.g., wetting), or by internal
kinetic friction, determined globally by the quality of contact surfaces
between grains. In contrast to rice-pile models of Refs.\
\cite{rice-transport,Lauritsen}, the present  models  are more
 appropriate for the evolution of landscape, in which a variety of
erosion mechanisms might be simultaneously active.

Two types of triggering mechanisms of landslides are
recognized  in the literature \cite{erosion,Romu,BT-gran}: (i) soil moisture,
which is controlled by rainfall and water level, and (ii) ground motion,
which influences slope variation. The local shear stress threshold may
depend on both slope angle and  soil properties. We assume that these
triggering mechanisms are dynamically correlated.  By wetting diffusion
probability is lowered and grains stick together, thus building up local
heights. However, when the {\it difference} between heights at neighboring
 sites exceeds certain limit, the slope mechanism becomes activated.

A simplified picture of the natural mechanisms of erosion is taken
into account
by combined relaxation rules for the  height  transport   on a
two-dimensional square lattice, as follows:
If at a site $(i,j)$ local height $h(i,j)\ge h_c$, then the site relaxes
{\it with probability }$p$ as $h(i,j)\to h(i,j)-2$ ;  $h(i+1,j_\pm) \to
h(i+1,j_\pm) +1$ . If for finite $p$ some of the local
slopes $\sigma _\pm (i,j)\equiv h(i,j)-h(i+1,j_\pm)\ge \sigma _c$, then the
site relaxes with probability {\it one}  by toppling one particle along each
unstable slope, repeatedly until all slopes are reduced below $\sigma _c$.
 Here $(i+1,j_\pm )$ are positions of two  downward neighbors of
the site $(i,j)$ on a square lattice oriented downward.

The system is driven by adding grain from the outside at a random site
along first
row  and the system is let to  relax according to the above rules.
Another grain is added when the relaxation process stops.
The internal time scale
is measured by the number of steps that the relaxation process  proceeds
on the lattice. At time step $t$=1 a site at first row topples
after added grain from outside.  According to the above relaxation rules,
one or two grains are toppled from that site, which then appear at
one or two downward neighboring sites. Therefore, mass flow is only down.
However,
an instability may propagate to both downward and upward neighbors of
a toppled site (except for the sites on the first row, which have no
upward neighbors), thus triggering four new sites as candidates for
toppling per each just toppled site.
At one time step we update {\it in parallel} all  candidates for
toppling. This comprises the usual definition of the time step in cellular
automata models.

Since the system builds up unstable sites (with respect to probabilistic
CH rule), the above dynamic rules need to be supplemented by an
additional rule, which makes the difference between two models.
In Model A, all sites that are visited by an avalanche at least once
are considered as candidates for toppling until the whole instability dies
off. In this way a propagating instability has an internal correlation
time $t_c$ which is determined by the dynamics itself.
In Model B, we set $t_c=1$. Therefore, only sites which are in the
neighborhood of active sites at time $t$ may be candidates for toppling in
the next time step $t+1$.
It should be stressed that, since an avalanche is
extended object, in both models there are many  sites which
topple simultaneously and which  are not neighbors in space.
In model B next toppled sites are neighbors only on time scale but not
in space, whereas in model A next toppled sites are not necessarily
 neighbors neither on temporal  nor spatial scale. However,
 all toppled sites are connected {\it within affected} area in space-time
 dimensions.
In both models particles are added from the outside
only on a random site at the first row and leave the system at lower (open)
boundary.  The mass transport is unidirectional (down).
However, since the above rules allow an instability to propagate both
forward and backward on a 2-dimensional lattice,
and evolve on an internal time scale, both models are
essentially $2+1$-dimensional, with extra dimension representing the
internal time scale. Differences in the additional relaxation
rule lead to different  emergent critical states, as explained
below.

In Fig.\ 1 two examples of large avalanches in model A (below) and model B
(top) are shown for values of the control parameter $p=p^\star$
at the edge of the  scaling region ($p^\star \sim 0.4$,
see Sec. IV for discussion).
In both models multiplicity of topplings at a site (larger number of topplings
is marked by  darker gray tone), is induced by the instability propagating
back and forth due to nonlocal relaxation rules. In model A number of
candidates for toppling at each time step is larger compared to model B,
due to internal correlation time typically $t_c\gg 1$,
leading to more efficient relaxation of unstable sites. On the other hand,
$t_c=1$ in model B enables building up numerous unstable sites (with
respect to CH rule) for low values of $p\sim p^\star$. Therefore  huge
avalanches  with perpendicular extent comparable  to the system size
(cf. Fig.\ 1 (top)) occur frequently, indicating that the anisotropy
of the relaxation events vanishes at $p^\star$.

In the limit $p=1$
both models reduce to the deterministic directed CH model introduced and
solved exactly by Dhar and Ramaswamy in Ref.\ \cite{DR}. In this limit
slopes are restricted to $\sigma _\pm (i,j) \le 1$, and thus
CS rule remains inactive.

 {\section{Model A: Multifractal scaling behavior of landslides}}

Correlation times $t_c>1$ in model A are motivated by varying
toppling conditions after an avalanche commenced, which represents
a  natural choice in the case of long relaxation times, such as
geological evolution of landslides. It has been shown that this type of
temporal disorder is a relevant perturbation both for the  evolution of
landslides \cite{BT-gran} and for  directed percolation processes
\cite{Jensen}. In this model each site develops an individual time
scale of activity, which then contributes to the whole event (avalanche).
As a consequence, the distribution of avalanche durations $P_A(t,L)$
in the scaling region exhibits  {\it  multifractal scaling} properties
when the system size $L$ is varied,  according to the expression:
\begin{equation}
P_A(t,L)=(L/L_0)^{\phi_t(\alpha _t)} ;~~~~~
\alpha _t\equiv \left(\ln{{t}\over{t_0}}\right)/\left(
\ln{{L}\over{L_0}}\right) .
\label{mfs}
\end{equation}
In Fig.\ 2 the distribution of duration of avalanches is shown for
$p=0.7$ and various lattice sizes. In the inset the spectral function
$\phi_t(\alpha _t)$ vs. $\alpha _t$ is determined by the scaling plot
according to Eq.\ (\ref{mfs}), with $t_0=1/4$ and $L_0=1/4$.
The integrated distribution of durations exhibits a power-law behavior as
$P(t)\sim t^{-(\tau _t -1)}$ in the entire region  $p^\star \le p< 1$,
with the $p$-dependent exponent $\theta \equiv \tau _t-1$, which is shown
in the inset to Fig.\ 3. Similar nonuniversality with decreasing scaling
exponents with the parameter $p$ are found for the  distributions of size
$D(s)\sim s^{-(\tau _s-1)}$, and mass
of avalanches  $D(n)\sim n^{-(\tau_n-1)}$ (see Ref.\ \cite{BT-gran} for
detailed analysis).
Slopes of various curves in the main Fig.\ 3
determine the dynamic exponent $z(p)$, which is also shown in the inset
to Fig.\ 3.
For values of the control parameter $p$ below a critical value $p^\star
\approx 0.4$ (see below) the critical steady states are no longer accessible
by the dynamics.

{\section{Model B: Nonuniversal scaling in granular piles}}

For finite correlation times, i.e., by setting $t_c=1$, avalanches
have in the average a  reduced number  of topplings per site, compared
with  model A for the  same value of the control parameter $p$.
This leads to a smaller incidence
of large avalanches, and thus to increase of the scaling exponents with
decreasing probability of toppling $p$.
In Fig.\ 4 the probability distribution of avalanche durations
 is shown for few values of $p$ in the scaling region. On the other hand,
for short correlation times the
balance between the CS and CH toppling mechanisms is altered:
By lowering $p$ the system builds up heights faster than in the case
of  model A, and
thus the CS mechanism becomes more effective, and eventually prevails at
the boundary of the scaling region at $p^\star$. We find numerically
that scaling behavior is lost at $p^\star\le 0.5$ \cite{LTU}.
The scaling behavior for $p^\star \le p < 1$ is characterized
by nonuniversal $p-$dependent scaling exponents (see inset to Fig.\ 5
and Ref.\ \cite{LTU}).

The scaling properties of the distribution of avalanche durations
are determined by using the following finite-size scaling form
\begin{equation}
P_B(t,L) \sim L^{-\theta z}{\cal{P}}(tL^{-z}) \ ,
\label{fss}
\end{equation}
where $\theta \equiv \tau _t-1$ as above, and  $z$ is the dynamic exponent,
 which also depends on $p$. The scaling plots of $ P_B(t,L)$ for various
values of $p$ in the scaling region and for three system sizes at each
value of $p$, are shown in the inset to Fig.\ 4. Similar scaling properties
are found for the distributions of size and length of avalanches (see
Ref.\ \cite{LTU} for detailed discussion). In addition to the temporal
distribution discussed above, here we also concentrate on the
 distribution of mass of avalanches, $P_B(n,L)$, satisfying the
scaling form $P_B(n,L) \sim n^{\tau _n-1}{\cal{Q}}(nL^{-D_n})$, where
mass $n$ of an avalanche is determined as total number of grains that slide
during one avalanche. In Fig.\ 5 the distribution of mass of avalanches
is shown for few different values of the parameter $p$ in the scaling
region.  In the inset to Fig.\ 5 we plot the
exponents $\theta (p)$ and  $\tau _n (p)-1$, for duration and mass of
avalanches, respectively, and the dynamic exponent $z(p)$, and
fractal dimension of mass $D_n(p)$ against $p$.
For $p\ge 0.5$ the following scaling relations are satisfied (cf. inset to
Fig.\ 5):   $(\tau _n-1)D_n  =z\theta =\alpha $, where
$\alpha \equiv \tau_\ell -1$ is the exponent of
length of avalanches, which is determined in Ref.\ \cite{LTU}.
The dynamic exponent $z$ which appears in the scaling form (\ref{fss})
can also be  determined from slopes of the curves  $<T>_\ell$ vs. $\ell$,
similar as we have determined it in model A.
 Obtained values are in a good agreement, within numerical error bars,
with those obtained from the scaling plots in Fig.\ 4.
Values of the exponents at $p=0.4$ are taken from the straight sections of
the lines representing distributions of duration and mass for smaller
system sizes $L\le 128$.
As indicated in the inset to Fig.\ 5, these values do not satisfy scaling
relations within error bars, indicating that $p=0.4$ is already beyond
the edge of the scaling region in model B (see discussion in Sec.\ IV).

{\section{Universal criticality  at the edge of the scaling region}}

When the control parameter $p$ is varied through a critical value $p^\star$
we find that the scaling behavior of the avalanche distributions is lost,
indicating that self-organized critical states are no longer accessible by the
dynamics. By numerical simulations of various distributions and applying
the appropriate scaling analysis it was shown that critical steady states
disappear  below $p=0.4$ in model A \cite{BT-gran}, and below
$p=0.5$ in model B  \cite{LTU}. Here we argue that  dynamic rules
with different correlation times $t_c$ in these models
lead to separate prevailing relaxation mechanisms  at the edge of the
scaling region, which lead to different values of
$p^\star $ and to two different universality classes of scaling
behavior.
In particular, in model A we find that the scaling exponents of
large avalanches $\theta (p^\star)$, $\tau _s(p^\star)$, etc.,
are close  to the universality class of parity-conserving (PC) branching
and annihilating random walks \cite{BARW,Jensen_PC}, whereas in model B the
exponents at $p^\star $ reach the values of the
 mean-field SOC universality class.

Although the relaxation rules in both model A and model B are complex
interplay of the probabilistic critical height and deterministic
critical slope
rules, we may distinguish two basic type of local branching and annihilating
processes that take part to propagate an avalanche in these models.
 Propagation of an avalanche may stop at a site to which one or two particles
drop in time step $t$, in the following two cases: (1) One particle drop
will not continue if the site had height zero, that is ``annihilation''
 $A\to 0$ occurs with probability $1-\rho $, where $\rho $ is the dynamically
changing probability that  a site has height $h\ge 1$; (2) When two
particles drop to a site at time $t$, the avalanche may not proceed
if the diffusion probability $p$ is too low, i.e., $A+A\to 0$ occurs
with probability $1-p$. Note that since number of particles is conserved
by the processes in the interior of the pile, ``annihilation'' means
accumulation of particles at a site, which thus will take part in future
events, in contrast to the case of chemical reactions, where particles
are extinct.
 When the conditions for toppling are fulfilled, propagation of an avalanche
represents a branching process which consists of two steps.
A toppled site at time $t$ transfers two particles forward, however,
the instability is transferred to its four neighbors,
but  the site itself can not topple in the next time step.
Toppling of an isolated site away from the open boundaries
by the critical height (CH) rule  makes
four neighboring sites as candidates for toppling in the next time
step, and if these four sites topple, they make nine new candidates
for toppling etc, along the chain $1\to 4\to 9\to 16 \to 25, \cdots $.
Since each toppled  site, both initial and  triggered sites, topple by two
particles, in CH mechanism,
this  toppling chain represents a reaction  $A\to (m+1)A$ with
odd number of offsprings $m=3,5,7,9, \cdots $ per each initial particle .
The same conclusion is true for  topplings by the critical slope (CS)
rule  with two simultaneously unstable slopes.
 If, however, a site topple by critical slope (CS) rule by dropping one
particle along one unstable slope, it will  trigger three neighboring sites
to topple by CS mechanism, and another toppling chain occurs as
$1\to 3\to 7\to \cdots $,  i.e.,  $m=2, 4, 6, \cdots  $ offsprings per
initial particle.

Diffusion limited branching annihilating random walks (BARW)
 have been studied by field-theory methods (for a  recent review see
\cite{Uwe} and references therein).
It has been recognized that $d_c=2$ is the upper
critical space dimension, and that BARW with even number of offsprings
 in $d=1$ belong to PC  universality class, whereas the directed percolation
(DP) universality class was found in the case of odd number of offsprings.
Two examples of dynamical processes in $1+1$
dimension belonging to  PC  universality class have been studied
numerically in Refs.\ \cite{PC1} and \cite{PC2}.
It should be stressed that  in contrast to BARW and DP processes,
the  present models A and B are dynamical, and thus the propagation
rules apply statistically and  depend on the history of the state of
the system.  Recently  an analogy  between the directed
percolation and stochastic dynamical model with critical height rules
has been discussed in Ref. \cite{TD}.

In model B, short correlation time makes the relaxation at a site less
effective with decreasing $p$, and thus efficient
building-up of heights occurs,  leading eventually to
$\rho (p^\star )\approx 1$. (A transverse section of the pile in model B
at $p^\star $ is given in  Fig.\ 6 (bottom)). Therefore, when
 an instability starts, it may trigger a mixture  of branching
processes described above, making
an instability transferred back and forth on two-dimensional lattice
and evolving in time. Fractal dimension associated with
the mass of avalanches at the edge of scaling region
was found to be $D_n(p^\star )\approx 2$ (see inset to Fig.\ 5).
Thus an avalanche appears to be compact in  $2$-dimensional
space and,  since next toppled sites are neighbors in
time ($t_c=1$), it represents a  connected object in  $2+1$ dimensions.
Starting an instability in {\it full} lattice, i.e., with
no threshold condition, will trigger an avalanche which propagates as
a directed percolation cluster
in 3-dimensions, until eventually too many sites will have heights
zero and the avalanche will stop. Thus $p^\star$ should coincide with
 the site-directed percolation threshold on simple cubic lattice
$ p_c^{SDP}=0.435$\cite{Grassberger}.
Mass of avalanche is defined as the number of particles that slide during
an avalanche, and thus it is equivalent to number of branchings.
Therefore, since for   $p=p^\star$ the  effective  dimension $D_n(p)$
 reaches the  upper critical dimension  of BARW, we  may expect
mean-field universality class for the scaling behavior of  avalanches.
Our numerical results listed in  Table\ 1 confirm this conclusion.
A schematic phase diagram is shown in Fig.\ 7.

The situation is different in model A (cf. Fig.\ 1), where decreasing
 diffusion probability $p$ an avalanche is either extinct quickly
(short avalanches), or lives much longer (large avalanches) with
large separation times \cite{BT-gran}.
 In turn, this leads to the efficient topplings at each affected site
due to many attempts within  correlation time $t_c\gg 1$.  In the resulting
steady state   most of the sites have heights $h< h_c$  (cf. Fig.\ 6 (top)).
Therefore, only toppling by CH rule takes place and threshold condition is
still active, in contrast to model B. In model A a toppled site $(i,j)$
at time $t$ may
trigger  topplings at time $t+1$ at  three neighboring sites, since
the site toppled at $t-1$ time step
will  not fulfill the threshold condition ($h\ge 2$) at time $t+1$.
It turns that among three neighbors less than two sites
topple  in the average, therefore leading to  a chain
of toppled sites with few branches, which is embedded in $2+1$
dimensional space-time. However, affected  sites which do not
topple due to low probability $p$ at first attempt may topple in
later time steps before the
avalanche dies off, thus starting a new chain. The avalanche
is made of set of such chains, and  has the fractal dimension
$D_n=1.48$. We believe that this effectively low dimensional BARW
process, although it takes part in  $2+1$ dimensional space-time, is
the reason for PC universality class in model A.  Another  reason
for the occurrence of PC universality class in reaction-diffusion processes
might be the existence of more than one symmetric absorbing states,
as discussed in \cite{PC1} and \cite{Monetti}.
The process is reminiscent to
bond-directed percolation in 3-dimensional simple cubic lattice, thus
we also expect that $p^\star \le p_c^{BDP}=0.382$ \cite{Grassberger}.
In the phase diagram in Fig.\ 7 phase boundaries for model A (dashed lines)
separate reactive phase from the critical and noncritical absorbing phases.

In the phase diagram in Fig.\ 7 phase boundaries for model A (dashed lines)
separate nonconducting phase from the conducting critical and noncritical
phases.
In  model B the noncritical conducting phase exists only along
the line $\rho =1$ below MF point, and a finite slope occurs via a phase
transition at SB point  \cite{LTU}. On the other hand, in model A our
results suggest that  noncritical steady states occupy a finite region
close to the right corner, and that a finite slope occurs asymptotically
at $p=0$, $\rho =1$. Further analysis is necessary in order to find
precise location of the PC point and the nature of phase transition between
critical and noncritical conducting  states. Along the phase boundaries
between the points PC and DR, and between MF and DR, we have the
nonuniversal criticality of model A and model B,  respectively,
 discussed in the present work. The point marked by DR at  $\rho =0.5$, $p=1$
corresponds to the universal SOC of Dhar-Ramaswamy model.

Sets of the exponents for $p=0.4$ are summarized in Table\ I.
Exponents in the model B at this value of $p$
are estimated from the straight sections  of lines in the subcritical
region for smaller lattice size $L=128$.
Value of the exponent $\tau _s$ is taken from  Ref.\ \cite{LTU}.
For comparison, shown are also the numerical values of the exponents
for PC universality class, from Ref.\ \cite{Jensen_PC}, and mean-field
self-organized criticality exponents, from Ref.\ \cite{TR}.
Note that our  exponent $\theta $ corresponds to the survival
probability distribution exponent
$\delta $ in Ref.\ \cite{Jensen_PC}, and  $z\equiv 2\nu_\bot /\nu_\|$
and that the scaling relation $\tau _s -1=\theta /(\theta +1)$ holds.
The exponents $\tau _n$ for mass of avalanche and roughness
exponent
$\chi $ are unique for granular piles, and can not be defined in
models of chemical reactions or damage spreading, considered
in Refs.\ \cite{PC1,PC2}.
 We estimate roughness exponent $\chi $ from the contour  of several
transverse  sections of the pile (two examples are given  Fig.\ 6).
For instance,
by  using box counting method we find the fractal dimension of the
contour curve  in model B, as $d_f=1.179-1.183$, and using  the
expression  $\chi =d_f-1$ leads to the value listed in Table\ 1.

{\section{Conclusions}}

We have shown that sandpile automata with mixed relaxation rules of
stochastic diffusion and deterministic branching processes are capable
of describing nonunuiversality of the self-organized critical states and
a loss of scaling at a critical value of the control parameter, in a
qualitative agreement with observed behavior in natural and laboratory
granular flow. Differences in the relaxation rules due to internal correlation
time lead to distinct dynamic critical states. In particular, unlimited
 (within lifetime of an  avalanche) correlation time $t_c$ in model A
leads to a multifractal scaling behavior and scaling
 exponents of large avalanches  {\it decrease} with decreasing values
of the control parameter $p$. On the other hand,
finite correlation time $t_c=1$ in model B leads to
{\it increase} of the scaling exponents with decreasing $p$,
 and to finite-size scaling properties of avalanches
in the entire scaling  region $p^\star \le p<1$.
 At the edge of the critical region at $p^\star$, dominating relaxation
mechanisms of modulo two conserving branching processes and effectively
low dimensionality of the  processes, lead to criticality
in the parity conserving universality class in model A. In model B
building up of a global slope appears to be dominant on top of the
above branching  processes, which thus appear to have the
effective dimension which exceeds the upper critical dimension
of BARW,  and thus  mean-field scaling exponents.
It should be stressed that the numerical
values of the exponents listed in Table\ I prove the closeness of
these universality classes within numerical error bars, which we estimate
as 0.03. Value of the exponent $\tau _n=1.66$ in mean-field models is
known only numerically \cite{TR}, whereas in branching processes,
which are equivalent to sandpiles with a fixed number of grains per
toppled site, there is the equality $\tau _n=\tau_s=3/2$.
 Study of the details of the  dynamic
phase transition in these models, e.g., in terms of the order parameter
and its fluctuations,  is left out of the present paper
(see, Ref.\ \cite{LTU} for appearance of finite slope at SB point
in model B).
 However, due to  scaling relations among various exponents  at the
transition, the observed different  universality classes of avalanche
statistics at $p^\star $ indicate that the exponents of the order
parameter $\beta $ and correlation length $\nu_\|$ should belong to
two distinct universality classes
of the dynamic phase transitions in these models.
Our results suggest that although basic relaxation rules in laboratory
granular piles and natural landslides might be the same, details of
actual implementation of these rules such as variation of control
parameter {\it during} the course of an avalanche might lead to
entirely different critical states.

\bigskip

\section*{Acknowledgments}

This work was supported by the Ministry of Science and Technology of the
Republic of Slovenia. I would like to thank Uwe T\"auber
 for helpful discussions.

\narrowtext
 \begin{table}[ht]
 \label{Table1}
 {\caption{Critical exponents at $p=0.4$ in models A and B, and for
parity-conserving (PC), and mean-field universality
classes (MF-SOC).}}
  \begin{center}
\begin{tabular}{|c|c c c c|}%\hline
%\hline
 E|M& Model A& PC& Model B&MF-SOC \\
\hline\hline
$\theta $ & 0.25& 2/7& 0.78 &3/4\\
\hline
 $z$ & 1.13& 8/7 & 1.33& 4/3 \\
\hline
  $\tau_s-1$ & 0.21 & 2/9 & 0.68&0.66 \\
 \hline
$\tau_n-1$ &0.19&---&0.52& 1/2  \\
\hline
$\chi $  &  0.05&---&  0.18& ?  \\
%  \hline
\end{tabular}
\end{center}
%\end{table}

%\newpage

\narrowtext
\begin{figure}[thb]
\epsfxsize=84mm\epsffile[18 245 592 685]{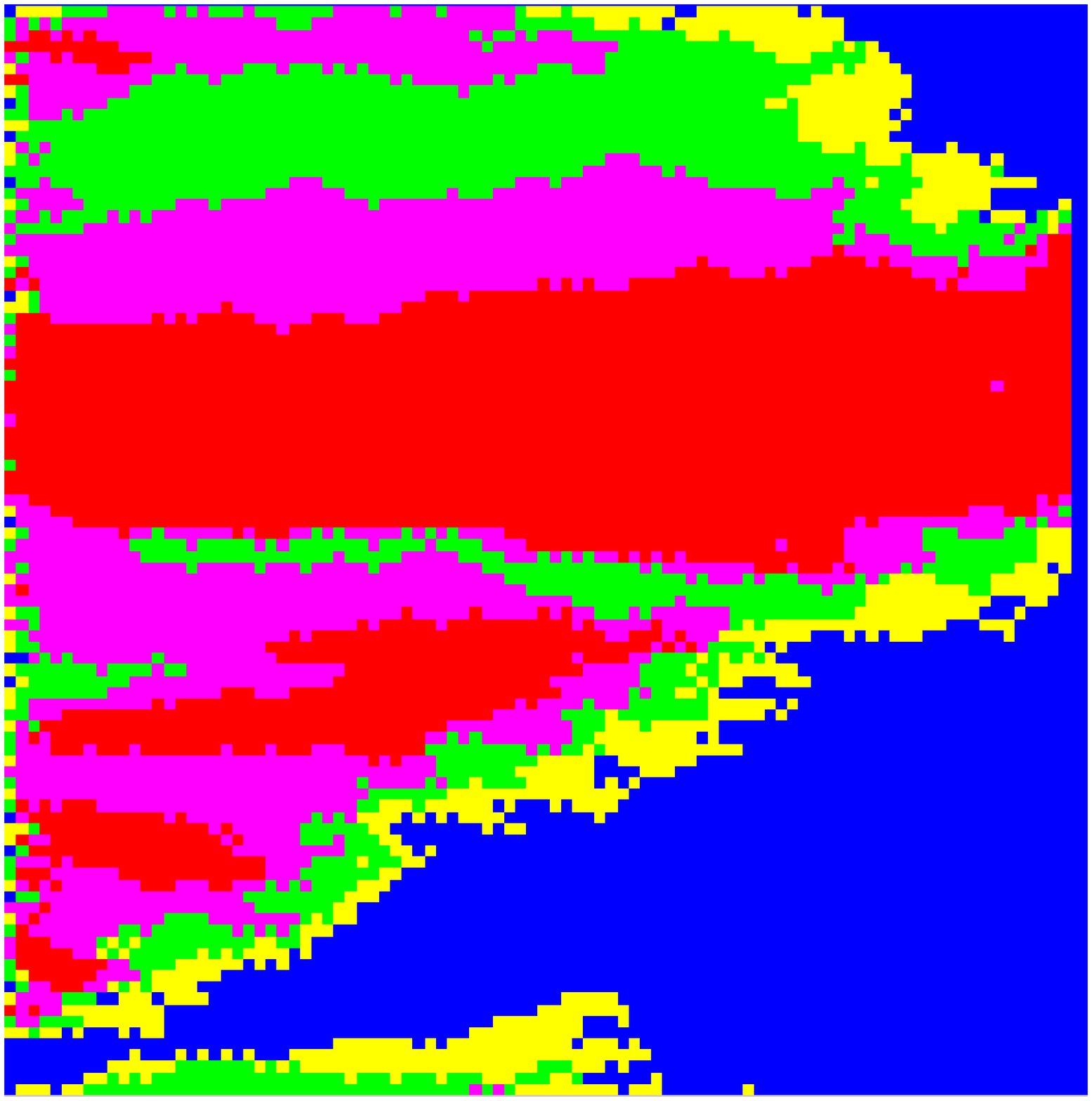}

\epsfxsize=84mm\epsffile[18 234 592 556]{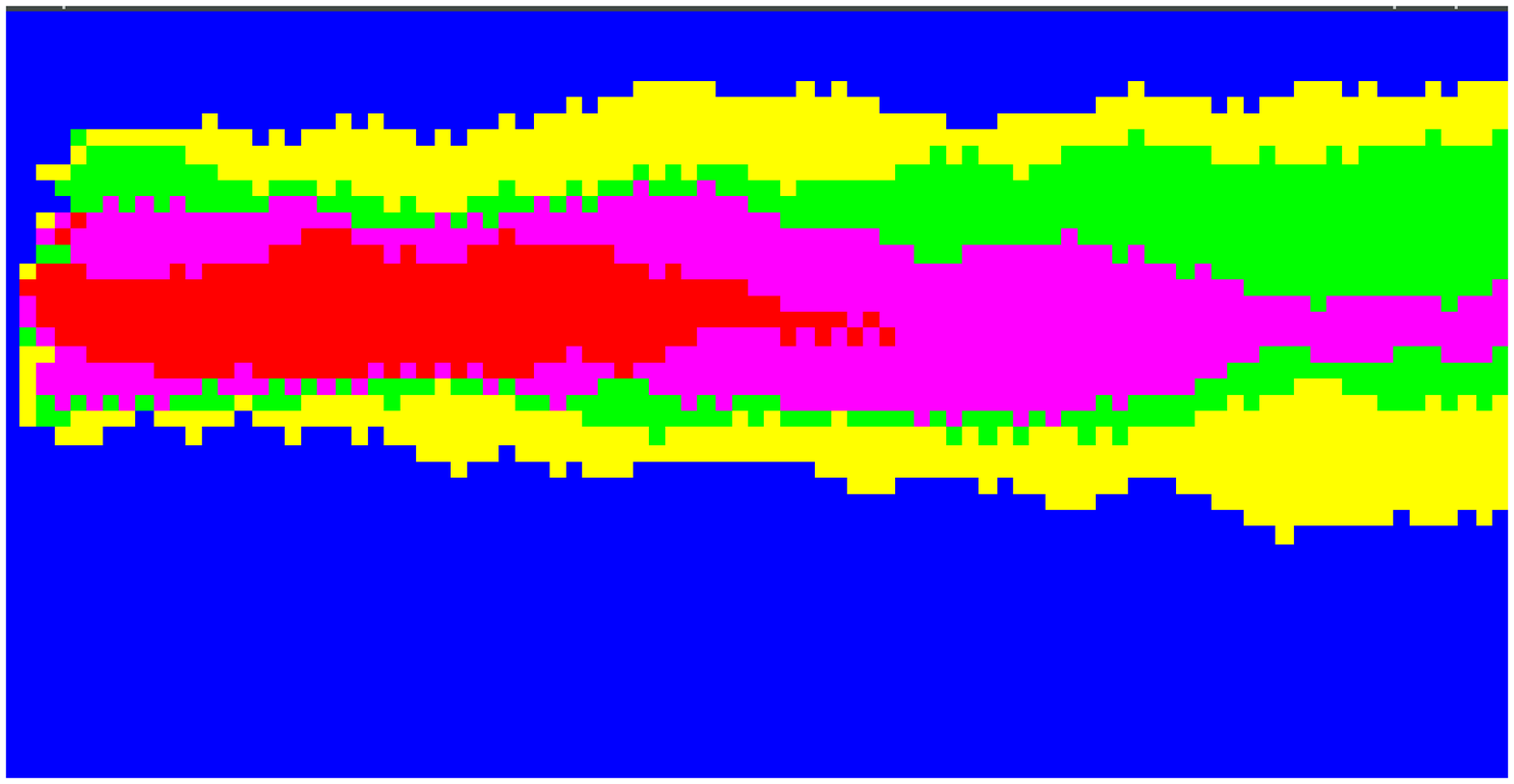}
\caption{\label{fig1} Two examples of large  avalanches running from left
to right at $p=p^\star$: in model A  (below) and
in model B (top). Multiple topplings up
to forth order are marked by different degree of gray color.}
\end{figure}

\begin{figure}[thb]
\epsfxsize=84mm\epsffile[34 70 565 562]{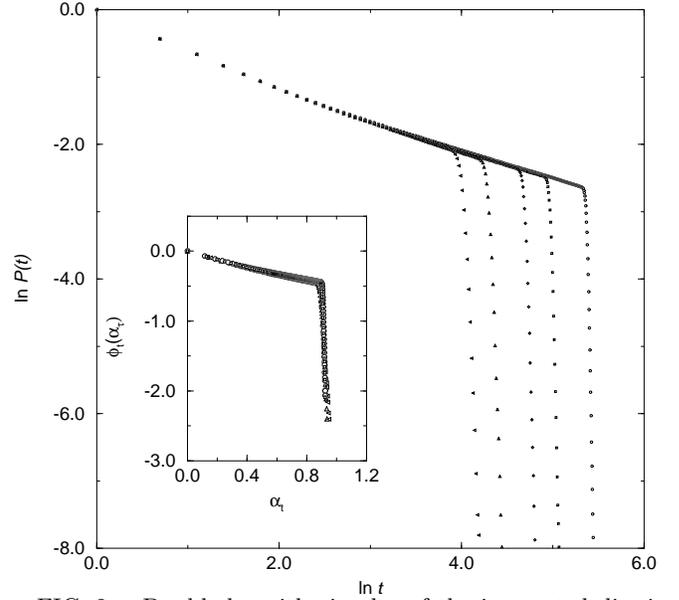}
\caption{\label{fig2} Double-logarithmic plot of the integrated
distribution $P(t)$ vs. $t$ for $p$=0.7 and for various
lattice sizes $L$=12, 24, 48, 96, and
192 (left to right)  in model A, obtained by open boundary conditions.
Inset: Multifractal spectral function $\phi _t(\alpha _t)$
vs. $\alpha _t$ . (Fig.\ 4 from Ref.\ [14]).}
\end{figure}

\begin{figure}[thb]
\epsfxsize=84mm\epsffile[67 68 582 578]{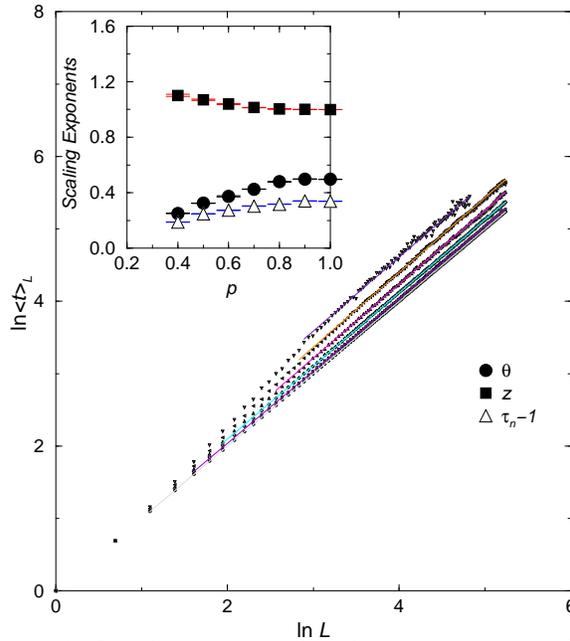}
\caption{\label{fig3} Average duration time of avalanches
of  selected length $<t>_L$ vs. length $L$  for various
values of $p$=1, 0.7, 0.6, 0.5, and 0.4 (bottom to top) in model A.
Maximal length equals the system size $L_{max}=$ 192, except for $p=$0.4,
where $L_{max}=$128. Inset: Scaling exponents $\theta \equiv \tau _t-1 $,
$z$, and $\tau _n -1$, defined in the text, vs. $p$ in the scaling region. }
\end{figure}

\begin{figure}[thb]
\epsfxsize=84mm\epsffile[40 68 582 580]{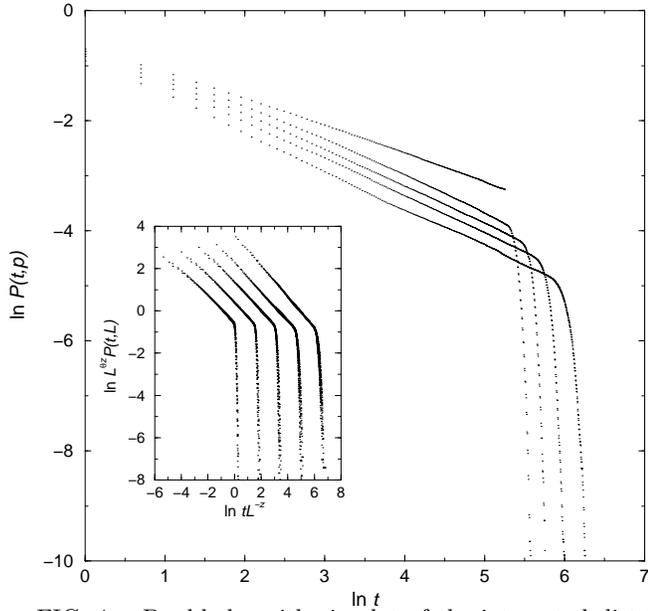}
\caption{\label{fig4} Double-logarithmic plot of the integrated
distribution $P(t)$ vs. $t$ for various values of the parameter
 $p$=1, 0.8, 0.7, 0.6, and 0.5 (top to bottom) and system size $L=$192
in model B.  Inset:
Finite size scaling plots for the same  values of $p$ as in the
main Figure (left to right). For each plot three different lattice sizes
 $L=$48, 96, and 192 are used.  Plots for different values of $p$
are shifted to the right for easier display.}
\end{figure}

\begin{figure}[thb]
\epsfxsize=84mm\epsffile[40 68 523 578]{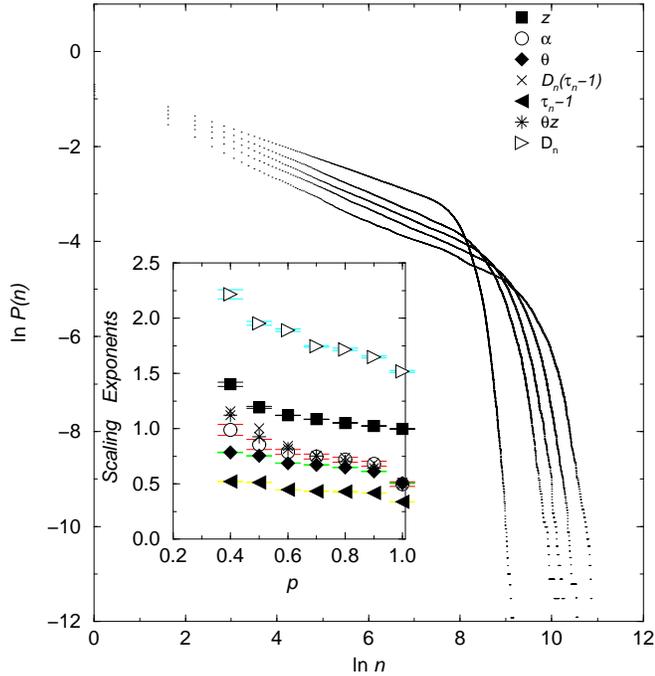}
\caption{\label{fig5} Double-logarithmic plot of the integrated
distribution of mass of avalanche $D(n)$ vs. mass $n$ for $L$=192
and for $p$=1, 0.8, 0.7, 0.6, and 0.5 (top to bottom) in model B.
Inset: Scaling exponents $\alpha \equiv \tau_\ell -1$,
$\theta \equiv \tau _t-1$, and $\tau _n-1$, fractal dimensions
 $z$ and $D_n$, and products $D_n(\tau_n-1)=z\theta $ plotted
against $p$ in the scaling region (see text).}
\end{figure}

\begin{figure}[thb]
\epsfxsize=84mm\epsffile[54 65 582 580]{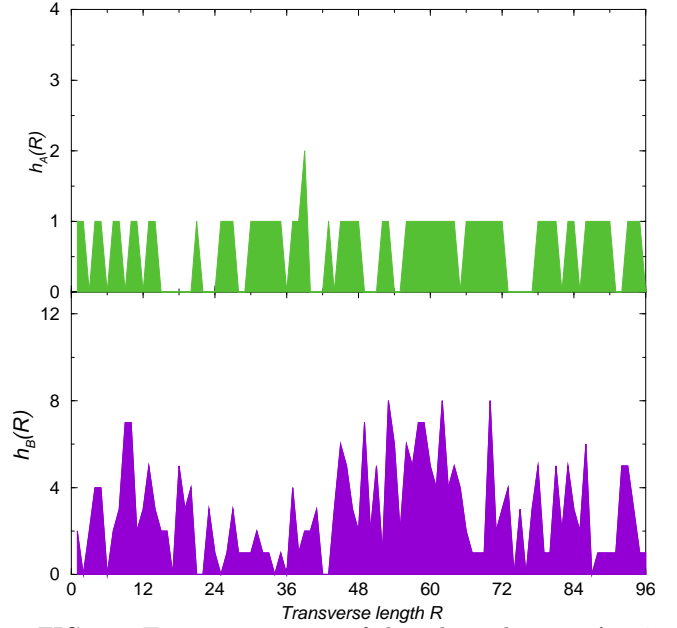}
\caption{\label{fig6} Transverse section of the pile at distance
$\ell =48$ from the top of pile for $p=0.435$ in model B (below), and
for $p=0.382$ in model A (top).}
\end{figure}

\begin{figure}[thb]
\epsfxsize=84mm\epsffile[51 90 480 461]{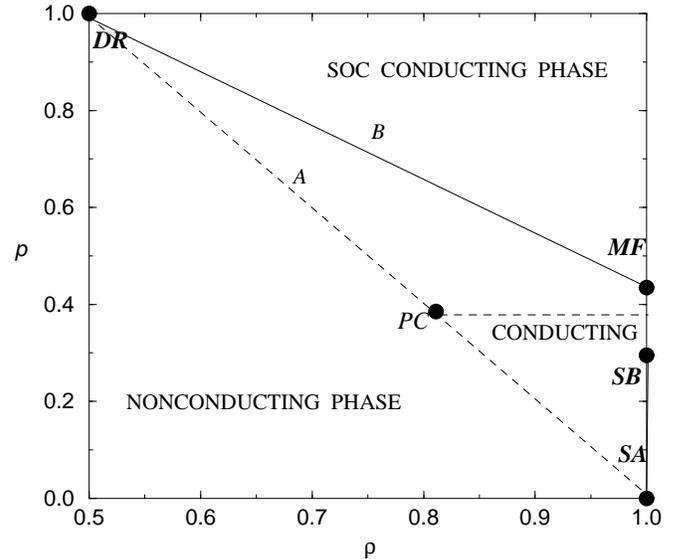}
\caption{\label{fig7} Schematic phase diagram in $(\rho ,p)$ plane.
Phase boundaries for model A (dashed lines) and for model B (solid lines).
}
\end{figure}
\end{table}
\end{multicols}
%\end{table}
\end{document}